# The photospheric origin of the Yonetoku relation in gamma-ray bursts

Hirotaka Ito[1,2], Jin Matsumoto[3], Shigehiro Nagataki[1,2,4], Donald C. Warren[1], Maxim V. Barkov[5,1] & Daisuke Yonetoku[6]

[1]Astrophysical Big Bang Laboratory, RIKEN, Saitama 351-0198, Japan

[2]Interdisciplinary Theoretical Science (iTHES) Research Group, RIKEN, Wako, Saitama 351-0198, Japan

[3]Department of Applied Mathematics, The University of Leeds, Leeds LS2 9GT, UK

[4]Interdisciplinary Theoretical & Mathematical Science Program (iTHEMS), RIKEN, Saitama 351-0198, Japan

[5]Department of Physics and Astronomy, Purdue University, 525 Northwestern Avenue, West Lafayette, IN 47907-2036, USA

[6]College of Science and Engineering, School of Mathematics and Physics, Kanazawa University, Kakuma, Kanazawa, Ishikawa 920-1192, Japan

**Long duration gamma-ray bursts (GRBs), the brightest events since the Big Bang itself, are believed to originate in an ultra-relativistic jet breaking out from a massive stellar envelope[1-2]. Despite decades of study, there is still no consensus on their emission mechanism. One unresolved question is the origin of the tight correlation between the spectral peak energy $E_p$ and peak luminosity $L_p$ discovered in observations. This "Yonetoku relation" [3,4] is the tightest correlation found in the properties of the prompt phase of GRB emission, providing the best diagnostic for the radiation mechanism. Here we present 3D hydrodynamical simulations, and post-process radiation transfer calculations, of photospheric emission from a relativistic jet. Our simulations reproduce the Yonetoku relation as a natural consequence of viewing angle. Although jet dynamics depend sensitively on luminosity, the $E_p$-$L_p$ correlation holds regardless. This result strongly suggests that photospheric emission is the dominant component in the prompt phase of GRBs.**

So far, no theoretical work has provided a fully consistent explanation for the origin of the Yonetoku relation. Both the well-studied internal shock model[5] and the more recent magnetic reconnection model[6] lack the ability to make firm predictions about the resulting emission properties, since they invoke non-thermal plasma physics with large uncertainties. Too many parameters (e.g., particle acceleration efficiency and magnetization) do not have a strong constraint but must be specified to evaluate the non-thermal emission. As a result, in order to reproduce the observed correlation, one needs to assume that there is a self-regulation among the imposed parameters[7]. However, it is not obvious why such a self-regulation should be satisfied across bursts.

In addition, models that invoke optically thin synchrotron emission also face problems in reproducing the spectrum (hard spectral slopes[8] and sharp spectral

peak[9]) in a non-negligible fraction of GRBs. These problems arise from the fundamental physics of synchrotron emission and so cannot be explained within this framework.

The above difficulties have led recent theoretical and observational studies to consider photospheric emission (photons released from a relativistic jet during the transition from optically thick to thin states) as a promising alternative scenario[10-24]. This model predicts the emergence of quasi-thermal radiation and can reproduce those spectral shapes that are incompatible with synchrotron theory.

Another strong advantage of the photospheric model is that it does not require a large number of uncertain parameters, since it is based on thermal processes. Indeed, many studies have discussed the origin of the relation based on photospheric emission. However, these analyses adopted oversimplified jet dynamics (e.g. steady spherical flow)[12,13] and/or crude assumptions for radiation processes[14,15]. More sophisticated study is necessary to firmly connect photospheric emission to the Yonetoku relation. We do so, robustly, here.

For an accurate analysis of photospheric emission, the jet evolution and accompanying photons must be followed from their origin, deep within the star, to the point where photons fully decouple from the jet. This requires both relativistic hydrodynamics and full radiation transfer. To capture all the essential features, the calculation needs to cover a large range in time and space, and must be performed in three dimensions (3D). We have previously reported on such a calculation[16], which was followed by another group[17,18] in 2D. However, these studies were only able to evaluate the emission at small viewing angles $\theta_{obs}$. High latitude ($\theta_{obs} \gtrsim 4°$) emission lacked accuracy since the calculation domains ($\lesssim 10^{13}$ cm) were not sufficient for the photons to decouple from the fluid in the jet. Moreover, the studies explored only a small part of the parameter space, so it was unclear how emission depends on the intrinsic properties of the jet.

To examine these issues, we perform large scale 3D relativistic hydrodynamical simulations of jets breaking out of a massive stellar envelope[25], followed by a post-process radiation transfer calculation in 3D. This procedure is well tested[16], but to achieve full decoupling of photons from the jet we extend the calculation domain by a factor of ~20 in space and time compared to our previous study. Moreover, we perform three sets of simulations to cover a wide range of model parameters. In each simulation, a jet with a different kinetic power is considered: $L_j = 10^{49}$, $10^{50}$ and $10^{51}$ erg/s (see the Methods supplement for details of our numerical setup).

Fig. 1 shows an image of our hydrodynamical simulation for the $L_j = 10^{50}$ erg/s model. Interaction with the stellar envelope, and the resultant formation of collimation shocks, most strongly influence the jet dynamics. Although this qualitative feature is common among the three models, it is most pronounced in the model with lowest jet power, since higher-power jets can penetrate the stellar envelope with less interaction. As a result, wobbling and complex structures are found throughout the outflow for the $10^{49}$ erg/s model. In the other two cases, only the portion nearest the jet head shows such features; the jet maintains a steady laminar structure below.

The resulting emission is summarized in Fig. 2. Models with higher jet power tend to show higher luminosity and spectral peak energy. This is mainly due to the larger energy budget for emission, and the higher overall temperature of the jet, as the jet power and Lorentz factor increase.

In the light curves, notable time variability arises due to the structure developed via jet-stellar interactions. For an observer with small viewing angle $\theta_{obs}$, steady emission is observed at later phases, since the inner region with laminar structure becomes visible. The fact that this feature is not observed in GRBs suggests that the central engines of these events are either not extremely luminous or not steadily luminous.

Regarding the spectra, we find non-thermal features compatible with observations, even though only thermal photons are injected in the current work. The broadening from a thermal spectrum at energies below and above the spectral peak is mainly caused by the multi-temperature and bulk Comptonization effects, respectively, which are induced by the global structure of the jet[13,19]. We note, however, that an accurate evaluation of the spectral shape requires higher spatial resolution[16,23]. Moreover, if present, non-thermal particles arising from internal dissipation[20-22] may also contribute to spectral broadening. Note that such dissipation does not affect the average energy of photons as long as the generated heat is smaller than the thermal energy. In the present study, we focus on the overall properties, such as spectral peak energy $E_p$ and peak luminosity $L_p$, that are largely unaffected by such ingredients.

A comparison of the Yonetoku relation with our results is shown in Fig. 3. We plot $E_p$ and $L_p$ sampled from the entire duration ~100s (roughly comparable to the duration of jet injection), but we also include the cases where only emission up to a certain duration (20, 40 and 60 s) is considered. This is intended to mimic bursts originating from shorter jet activity, since long GRBs have diversity in their durations. Since the early phase of the emission is nearly identical to the entire emission arising from shorter jet injection[24], we consider this simple change justified.

The lateral structure of the jet, developed during propagation through the stellar envelope, leads to a strong dependence on the viewing angle. Since the region near the jet axis has the highest Lorentz factor and temperature, one expects higher luminosity and spectral peak energy at smaller $\theta_{obs}$. This sequence produces a continuous correlation between $E_p$ and $L_p$ that spans several orders of magnitude. Though the distributions of $E_p$ and $L_p$ are shifted to higher values as the jet power increases, we find very similar behavior in all three models despite the variety in dynamics. All models reproduce the Yonetoku relation remarkably well. Since a wide range of jet power and duration is covered in our analysis, we stress that this is not the result of fine-tuning in our simulation setup but an inherent property of GRB photospheric emission.

We see some dependence on duration, as shorter durations tend to slightly shift the spectra to the softer side. This is because the region near the head of the jet is subject to baryon loading from the progenitor envelope, which pushes out the photosphere to larger distances, and therefore cools the radiation. Although this causes some dispersion in the correlation, all three models trace the Yonetoku relation regardless of duration. We conclude that the tight correlation is not affected by the duration of jet injection, or by the jet power. Simultaneously, the insensitivity to duration indicates that the correlation holds for any time interval of the emission, as in GRB observations[26,27] (see also methods).

Contrary to predictions in previous studies based on 1D models[12,13], the Yonetoku relation need not reflect diversity in the intrinsic properties of the jet. Instead, it naturally arises from the dependence of emission properties on the viewing angle. Bright, hard emission is observed on-axis, while soft, dim emission can be observed off-axis. Variation in the jet properties (e.g., power, Lorentz factor, and duration) appears as dispersion in $E_p$ and $L_p$ around the correlation curve, which also nicely reproduces the observed scatter.

Previous works that performed 2D hydrodynamical simulations[14,15] also claimed to reproduce the observed correlation between the spectral peak energy and the total radiated energy (the Amati relation)[28] through changing the viewing angle. However, their calculations did not include radiation transfer, which is essential for the evaluation of the emission. Indeed, recent simulations that do incorporate radiation transfer calculation show deviations from previous results[17,18]. We note again that the imposition of 2D axisymmetry and the limited calculation domain can cause inaccuracy. The former assumption induces error in the evaluation of emission, particularly along the jet axis because of the coordinate singularity, while the latter ingredient prevents robust predictions of off-axis emission. Our current study overcomes both issues and shows, for the first time, that the Yonetoku relation is an inherent feature of photospheric emission in GRBs.

While our results show a continuous sequence over three orders of magnitude in $E_p$, the observational data are limited at luminosities below ~$10^{50}$ erg/s due to difficulties detecting dim transients. We find excellent agreement at high luminosities (>$10^{50}$ erg/s) where the observations do not suffer from possible biases suggested in the literature[29]. Here the best fit curve of our simulation is given by $L_p = 10^{52.6} \times [E_p/355\text{keV}]^{1.67}$ erg/s, which is consistent with the observations (see Fig.3). On the other hand, at low luminosities where observations do not provide a strong constraint, we find a slight deviation of population away from the best fit curve of the Yonetoku relation towards higher peak energy $E_p$. Nevertheless, the simulation results also overlap with the only existing observational data at such low luminosities. This may be indicating that the correlation curve has a steeper slope at this luminosity range. We look forward to future observations falling in this part of the $E_p$-$L_p$ plane.

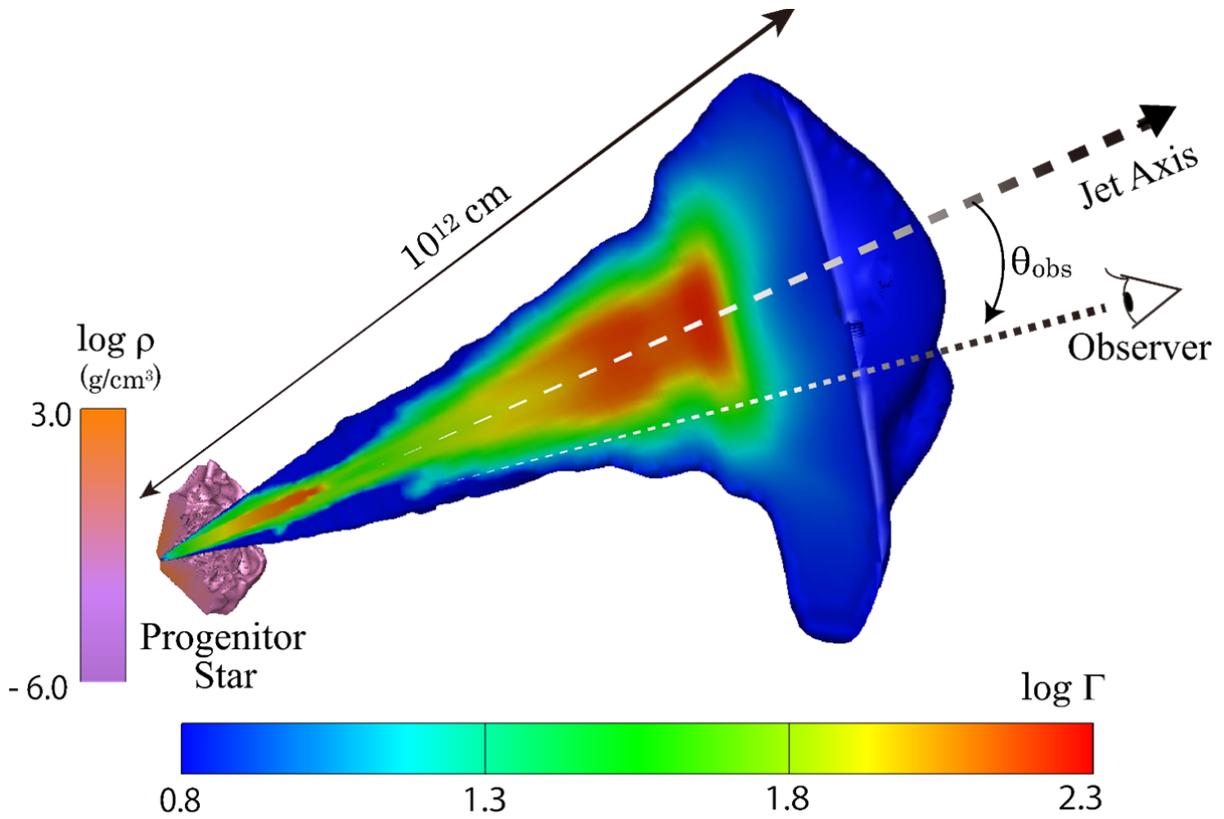

**Figure 1| Snapshot of the hydrodynamical simulation.** 3D profile with a 2D slice taken through the midplane of the simulation at a laboratory time t = 40 s for the model with jet power $L_j = 10^{50}$ erg/s. The profiles of the progenitor star and jet are visualized using color contours of mass density and Lorentz factor, respectively. Together with the simulation result, we also show the location of the jet axis (dashed arrow) and how we define the viewing angle $\theta_{obs}$ of an observer's line of sight (dotted line).

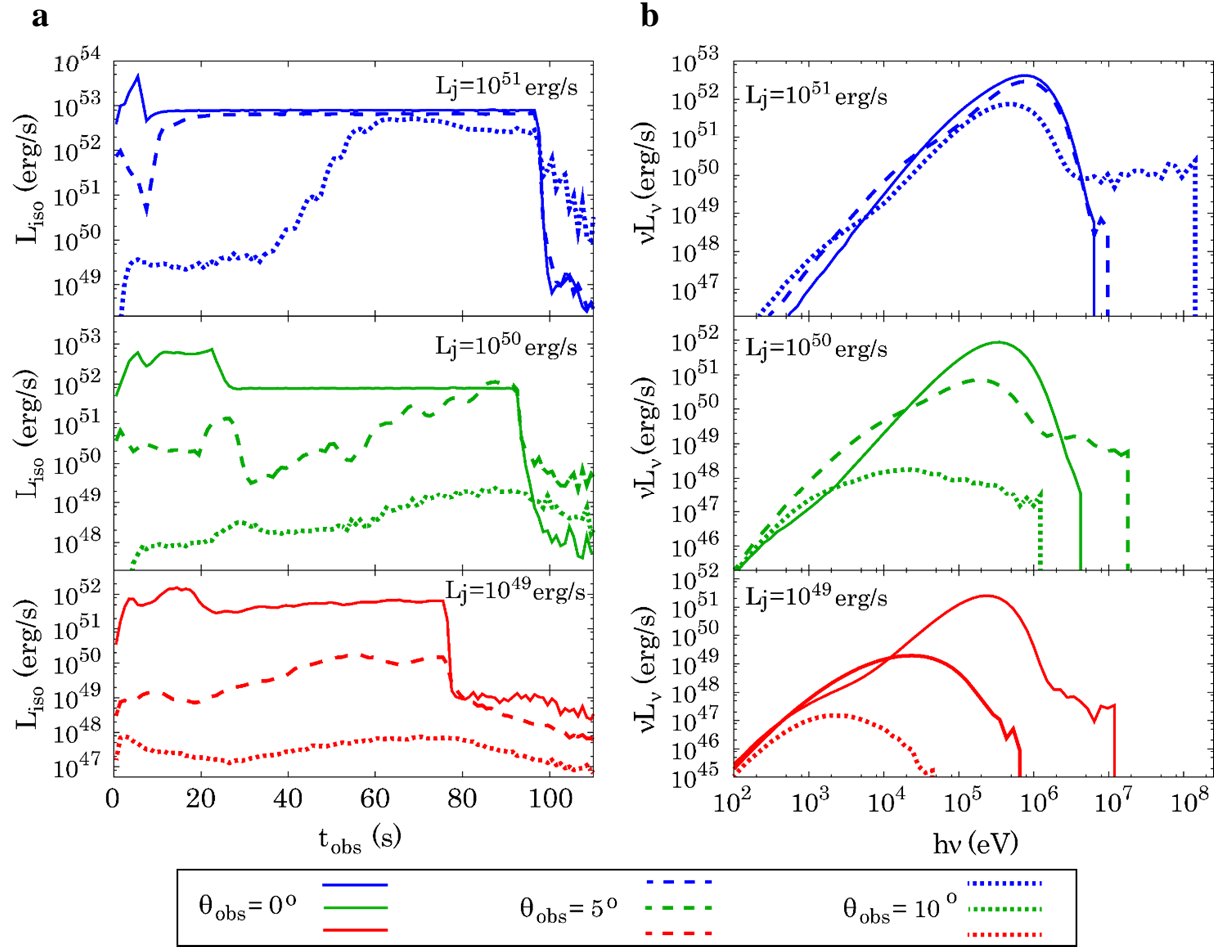

**Figure 2| Light curves and time integrated spectra. a,** Light curves up to observer time $t_{obs}$ = 110s. **b**, Time integrated spectra constructed by averaging over the duration $t_{dur}$ = 110s. Line color indicates jet power, with red, green and blue showing the cases of $L_j$ = $10^{49}$, $10^{50}$ and $10^{51}$ erg/s, respectively. For each model, we show results for three different viewing angles: $\theta_{obs}$ = 0° (solid line), 5° (dashed) and 10° (dotted). Note that, although high energy photons suffer from low statistics, this does not affect the evaluation of the overall luminosity or the spectral peak energy.

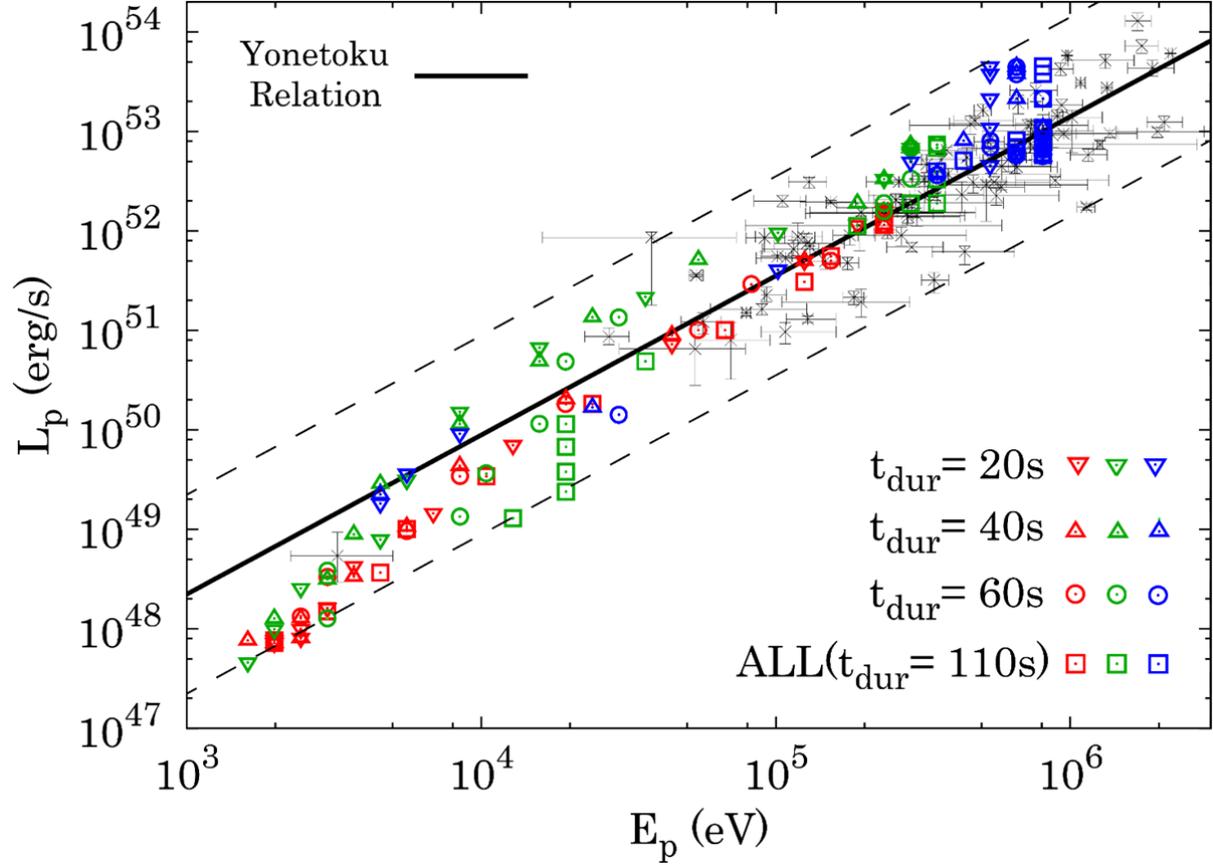

**Figure 3| Relation between spectral peak energy $E_p$ and peak luminosity $L_p$.** Results of our simulations are plotted with the observational data of 101 GRBs (gray points with error bars) and best fit curve of the Yonetoku relation (black solid line), $L_p = 10^{52.43\pm0.037} \times [E_p/355\text{keV}]^{1.60\pm0.082}$ erg/s, taken from the literature[4]. Two dashed lines located below and above the best fit curve show the 3-σ systematic error regions of the Yonetoku relation. Symbol color indicates model, with red, green and blue representing jet powers of $L_j = 10^{49}$, $10^{50}$ and $10^{51}$ erg/s, respectively. The inverted triangle, triangle, circle and square plot the results obtained by sampling the emission up to durations $t_{dur}$ = 20, 40, 60 and 110 s, respectively. The considered range of viewing angle is $0° \leq \theta_{obs} \leq 11°$, with 1° between successive points. Although the current results do not extend up to the bright end of the observed distribution, we expect that this population can be naturally produced by increasing the jet power and/or Lorentz factor, which shift the population toward higher energy and luminosity in the present study.

## METHODS

**Hydrodynamical simulation.** In order to evaluate the long-term evolution of a relativistic jet that penetrates a massive star, we have performed special relativistic hydrodynamical simulations in a three dimensional spherical coordinate system $(r,\theta,\phi)$. The numerical code is identical to that of our previous work[16]. The main difference in the numerical setup is that we have followed the evolution for a longer time scale (up to t = 6000s) and, therefore, larger spatial scale (up to r ~ $2\times10^{14}$ cm). Note that this is mandatory to ensure that the photons are fully decoupled from the jet even in at high latitudes[17,18]. In order to accomplish the calculation within a reasonable computational time, we have done six remappings in the radial direction. The corresponding time and radial spatial domain at each remapping are as follows:

**0th**   $0s \leq t \leq 130s$
     $10^{10}$ cm $\leq r \leq 8\times10^{12}$ cm

**1st**   $130s < t \leq 200s$
     $10^{11}$ cm $\leq r \leq 2.9\times10^{13}$ cm

**2nd**   $200s < t \leq 1000s$
     $10^{12}$ cm $\leq r \leq 6.1\times10^{13}$ cm

**3rd**   $1000s < t \leq 2000s$
     $8\times10^{12}$ cm $\leq r \leq 1\times10^{14}$ cm

**4th**   $2000s < t \leq 3500s$
     $2.9\times10^{13}$ cm $\leq r \leq 1.4\times10^{14}$ cm

**5th**   $3500s < t \leq 5000s$
     $6.1\times10^{13}$ cm $\leq r \leq 1.8\times10^{14}$ cm

**6th**   $5000s < t \leq 6000s$
     $10^{14}$ cm $\leq r \leq 2.2\times10^{14}$ cm

In each remapping process, we shift the position of the inner and outer boundaries to larger distances by discarding and adding the inner and outer grid zones, respectively, so that the entire structure of the propagating jet is contained within the calculation domain at any time. The two angular coordinates ($\theta$ and $\phi$) are fixed at $\pi/4 \leq \theta, \phi \leq 3\pi/4$ in all remapping. We impose a reflective boundary condition on the initial (0th) inner boundary in the radial direction except for the jet injection region, while an outflow (zero gradient) boundary condition is employed after the 1st remapping. The outflow boundary condition is also used for the outer boundary of the radial grid as well as all four boundaries along the side of the grid throughout the calculation.

As for the spatial resolution of the calculation, 280 uniformly spaced grid zones are used for $\theta$ and $\phi$, while 1260 non-uniform grid zones are employed for the r coordinate. The grid size in the radial direction increases with radius as $\Delta r = \Delta\theta\, r\, [1 + r/r_0]^{-1}$. In this equation, $\Delta\theta = \pi/560$ is the angular grid size and $r_0 = 2\times10^{13}$ cm is the reference position beyond which radial grid size asymptotically approaches a constant value. The suppression of increase in grid size at $r > r_0$ is introduced to maintain the radial resolution at a level where the overall jet structure can be resolved even at large radius, $r \sim 10^{14}$ cm.

In solving the special relativistic hydrodynamics, we use the numerical code developed by one of the authors[30] which employs a relativistic HLLC Riemann solver scheme. A MUSCL-type interpolation method is used to attain second-order accuracy in space, with second-order temporal accuracy using Runge–Kutta time integration. We assume an ideal gas equation of state, $p = (\gamma - 1)\,\rho\,\varepsilon$, with $p$, $\gamma = 4/3$, $\rho$ and $\varepsilon$ being pressure, specific heat ratio, rest mass density and specific internal energy, respectively.

Though they are irrelevant in the hydrodynamical evolution, the local temperature, $T$, and number density of electrons, $n_e$, must be specified for the calculation of radiation transfer. The temperature is determined from the pressure in assumption of a radiation dominated gas, namely, $p = a_{rad} T^4 /3$, where $a_{rad}$ is the radiation constant. The electron number density is determined from the mass density as $n_e = \rho/m_p$ under the assumption of full ionization, where $m_p$ is the rest mass of the proton.

As the initial condition of the simulation, we consider a massive progenitor star that is surrounded by a dilute ambient gas with a wind-like profile. The progenitor star is modeled as a Wolf-Rayet star with mass of ~14 solar mass at the presupernova stage, taken from model 16TI in the literature[25]. Beyond the radius of the stellar surface $R_* = 4 \times 10^{10}$ cm, where a sharp drop in density occurs, we continuously connect to the external dilute gas that has a decaying power-law profile given by $\rho = 1.7 \times 10^{-14} (r/R_*)^{-2}$ g / cm$^3$.

Given the initial condition, subsequent hydrodynamical evolution is governed by the jet that drills through the massive stellar envelope. In the current study, we carry out three sets of simulations that consider jets with different kinetic luminosities: $L_j = 10^{49}$, $10^{50}$ and $10^{51}$ erg/s. In all cases, the jet is continuously injected from the inner boundary of the initial (0th) grid (r = $10^{10}$ cm) with a half-opening angle and Lorentz factor given by $\theta_j = 5°$ and $\Gamma_i = 5$, respectively. While the initial specific heat ratio is fixed at $h_i = 100$ for the models with $L_j = 10^{49}$ and $10^{50}$ erg/s, a higher value $h_i = 180$ is adopted in the model with $L_j = 10^{51}$ erg/s. This means that the model with the highest jet power also reaches the highest terminal Lorentz factor, typically given by $\Gamma_i h_i$. The central axis of the jet is aligned to the radial direction at $\theta = \phi = \pi/2$. We suddenly stop the steady injection at t = 100 s and compute the evolution until the head of the jet reaches ~2×$10^{14}$ cm by utilizing the remapping described above.

**Radiation transfer calculation.** Radiation transfer in three dimensions is calculated using a Monte-Carlo method. As in the case of the hydrodynamical simulation, the method and setup is identical to our previous work[16], but longer time evolution is considered. By employing the output data of the hydrodynamical simulation as a background fluid, we directly track the trajectories of photon packets, which are an ensemble of multiple photons that have identical 4-momenta.

Initially, the photon packets are injected at the surface of a partial sphere, at a radius determined by the optical depth from an infinitely distant observer along the jet axis $\tau(r) = \int_r^\infty \Gamma n_e \sigma_T (1 - \beta \cos\theta_v)\, dr'$, where $\Gamma, \beta, \sigma_T$ and $\theta_v$ are the bulk Lorentz factor, 3-velocity normalized by the speed of light, Thomson cross section and angle between the line-of-sight (LOS) of the observer and velocity direction, respectively. Here we choose a value of $\tau = 100$ for the injection radius. We note that our results depend only weakly on injection radius. Any deviation from thermalized distributions at greater optical depths does not survive to the photosphere[22]. The solid angle of the photon injection surface is the region with bulk Lorentz factor larger than 1.5, in order to focus on the photons emerging from relativistic outflow.

At the given surface, photons are injected with an intensity of the black body emission of local temperature. Due to relativistic effects, intensity at a frequency $\nu$ is evaluated as $I_\nu = [\Gamma(1 - \beta \cos\theta_\nu)]^{-3} B_{\nu'}(T)$, where $B_{\nu'}(T) = 2h\nu'^3 c^{-2}[\exp(h\nu')/kT - 1]^{-1}$ is the Planck function. Here $\nu' = \Gamma(1 - \beta \cos\theta_\nu)\nu$ is the frequency at the comoving frame, and $h$ and $k$ are the Planck constant and Boltzmann constant, respectively.

Based on the given intensity, our code initially distributes numerous photon packets at the injection surface. Then the packets undergo a large number of scatterings by electrons, and are tracked until they reach the outer boundary of the calculation domain of the final (6th) series of the remapping r ~ $2 \times 10^{14}$ cm. In our code, the distance between the scattering events is determined by drawing the corresponding optical depth $\delta\tau$. The probability for the selected optical depth to be in the range $[\tau, \tau + \delta\tau]$ is given by $\exp(-\delta\tau)d\tau$. For a given optical depth, the physical distance is computed by integrating the opacity $n_e \sigma_{KN}(1 - \beta\cos\theta_\nu)$ along the path of the photon over the time-evolving background fluid, where the total cross section for Compton scattering, $\sigma_{KN}$, fully takes into account the Klein-Nishina effect. At the scattering event, we first choose the 4-momentum of the electron which interacts with the photon, drawn from a Maxwell distribution of local temperature. Then we transform the 4-momentum of the photon to the rest frame of the electron and determine 4-momentum after the scattering based on a differential cross section for Compton scattering. Finally, we update the 4-momentum of the scattered photon by transforming it to the frame of a stationary observer. The local temperature and velocity at the scattering position is determined by linear interpolation from the surrounding grid centers.

**Light curves and spectral analysis.** By sampling the photon packets that have reached the outer boundary, we determine the properties of emission in the observer frame. For a given viewing angle, the light curve and spectrum are computed by collecting the photon packets that have propagation directions contained in a cone of half-opening angle $0.5^\circ$ around the LOS. The imposed opening angle is small enough to ensure convergence of our results. The time interval used to construct the light curve is 1 s, identical to that used in the observation to find the peak luminosity[3,4]. In constructing the time integrated spectra, we divide the energy range from $h\nu = 10$eV up to 10 GeV in 100 bins equally spaced in a logarithmic scale ($\nu_n/\nu_{n-1} = 1.23$). In the current study, we consider four choices for duration of the time integration: 20, 40, 60 and 110 s. For a given duration $t_{dur}$, the spectral peak energy $E_p$ is determined by specifying the frequency at which the corresponding time integrated spectra $\nu L_\nu$ show a peak, while the peak luminosity $L_p$ is determined by identifying the maximum luminosity in the light curves within the duration.

The total number of packets injected in each model is $7\times 10^8$. This is sufficiently large to attain statistical convergence of our results except for the very highest photon energies.

Since our calculation is performed in three dimensions, the jet is not axisymmetric. Hence, the emission depends not only on the viewing angle, but also on the azimuthal angle. However, the dependence is not strong. Dispersion in the values of $E_p$ and $L_p$ due to the azimuthal angle is within a factor of 3, and the results always reproduce the Yonetoku relation. Therefore, we only show the results for a fixed azimuthal angle as a representative case.

**On the validity of the numerical setup.** The main focus of the current study is on the spectral peak energy $E_p$ and peak luminosity $L_p$. One crucial ingredient that governs these quantities is the temperature of the outflow. In our calculation, we assume that a black body is realized throughout the flow in order to determine the temperature. It must be noted, however, that this prescription loses accuracy once dissipative heating takes place at regions with an optical depth $\tau \lesssim 10^5$. This is because photon production is too slow to achieve full thermal equilibrium, as shown in the literature[24]. Hence, assumption of black body overestimates photon number density in the presence of dissipation, which in turn leads underestimation of temperature.

In the three simulations performed in the current study, the black body assumption is valid at the injection radius $r = 10^{10}$ cm in all three models due to the high optical depth at the injection region ($\tau \sim 10^5$). However, since dissipative heating (via the formation of intense shock) takes place in the outflow during propagation, the photon distribution begins to fall out of thermal equilibrium at larger radii. Nevertheless, we emphasize that error caused by departure from black body because of shock dissipation is not large. We justify this claim below.

First, let us briefly summarize the hydrodynamical properties of the jet considered in our simulation. Our assumption in all three models is that we inject a radiation-dominant (i.e., internal energy of the radiation is larger than the rest mass energy density) outflow which has a potential to accelerate up to a bulk Lorentz factor of few 100s. Since the injection radius is located at $r = 10^{10}$ cm and the initial bulk Lorentz factor is 5, this means that the outflow continues to be radiation dominant at least up to the saturation radius $\sim 10^{12}$ cm for a simple radial adiabatic expansion. Note that the radiation dominant region extends to larger distances in the actual flow, since shock dissipation is present.

Shocks formed in the radiation dominant phase are considerably less efficient at heat generation (which increases of photon-to-baryon ratio $n_{ph} / n_b$ of black body radiation) than those in the matter dominant phase[23]. This is due to the fact that the flow upstream from the shock is already hot, and so the shock provides minimal additional heat in this case. As a result, our prescription does not lead to a large inaccuracy in the temperature estimation roughly up to the saturation radius $r \sim 10^{12}$ cm.

On the other hand, shocks can lead to some error in the temperature estimation at larger radii, after the flow becomes matter dominant. However, the optical depth in this region has decreased below the value that can sustain saturated Comptonization ($\tau \lesssim 100$ ;

"unsaturated Compton zone"[22]). In this region, photons cannot immediately respond to the rapid temperature change due to dissipative heating before they decouple from the jet and escape. Therefore, dissipation does not have a significant effect on the resulting $E_p$ and $L_p$. In other words, accuracy in tracking the rapid temperature change caused by dissipation is not crucial for an evaluation of these quantities. Moreover, we note that most shock heating occurs during propagation through the progenitor star, so only a small fraction of the jet matter is shock heated at these large radii. This fact further reduces the error caused by the shocks.

The above qualitative discussion explains why our prescription for the temperature does not induce a large inaccuracy in the evaluation of $E_p$ and $L_p$. Of course, further quantitative estimation is necessary to ensure that this claim is robust. For this purpose, we perform additional radiation transfer calculations. Here we employ the same three sets of hydrodynamical simulations as a background fluid, but we lift the assumption that a black body is realized throughout the outflow in determining the temperature. Instead we assume that, while photons at the base of the jet (r = $10^{10}$ cm) form a black body, the photon to baryon number ratio is conserved thereafter. With the local photon number density determined by the new prescription, the temperature can be calculated from the EOS by $p = n_{ph} k T$. In the absence of dissipative heating, this prescription coincides with the original one. However, once dissipation begins to play a role, it leads to a larger temperature. While the original prescription corresponds to the limit of efficient photon production (immediately leading to full thermal equilibrium), this is the limit of inefficient photon production. Since the true solution should be found in between the two cases, the difference in the resulting $E_p$ and $L_p$ represents the uncertainty caused by the assumption for the temperature.

As mentioned above, our new prescription tends to increase the temperature from the original calculation. However, rare regions with lower temperature also appears due to the entrainment of the external medium, which originally had much lower photon number density. These cases are not significant, and we again employ the prescription of black body since the lower temperature is unlikely to be physical.

Note that we also change initial condition of the thermal photons at the point of injection to be consistent with the new prescription. Namely, we set the temperature and number density of the thermal photons to coincide with the newly determined values.

The resulting $E_p$ and $L_p$ for the new calculation is shown in Supplementary Figure 1. As is apparent from a comparison with Fig. 3, no significant discrepancy is found between the two cases. Therefore, we can robustly conclude that our result is not affected by the ambiguity in the evaluation of temperature.

**On the time resolved analysis of $E_p$-$L_p$ correlation.** Although not as established as the correlation found among the bursts, there is an important indication in literature[26,27] that $E_p$-$L_p$ correlation also holds at any time interval of individual bursts. To see whether such tendency is also found in our calculation, we performed a time-resolved analysis of our results. Here, we have taken uniform time intervals of 10s and determined the spectral peak and peak luminosity within the each interval. The results are displayed in Supplementary Figure 2. As shown in the figure, we also find a good agreement with the correlation curve. Hence, our calculation supports the picture that $E_p$-$L_p$ relation is also satisfied within an individual burst.

References

1. MacFadyen, A. I, Woosley, S. E. Collapsars: Gamma-Ray Bursts and Explosions in ``Failed Supernovae". Astrophys. J. 524, 262-289 (1999)
2. Aloy, M. A., Müller, E., Ibáñez, J. M., Martí, J. M., MacFadyen, A. Relativistic Jets from Collapsars. Astrophys. J. 531, L119-L122 (2000)
3. Yonetoku, D., Murakami, T., Nakamura, T., Yamazaki, R., Inoue, A. K., Ioka, K. Gamma-Ray Burst Formation Rate Inferred from the Spectral Peak Energy-Peak Luminosity Relation. Astrophys. J. 609, 935-951 (2004)
4. Yonetoku D., Murakami, T., Tsutsui. R., Nakamura. T., Morihara. Y., Takahashi. T. Possible Origins of Dispersion of the Peak Energy-Brightness Correlations of Gamma-Ray Bursts. Publ. Astron. Soc. Jpn. 62. 1495-1507 (2010)
5. Rees, M. J., Mészáros, P. Unsteady outflow models for cosmological gamma-ray bursts. Astrophys. J. 430, L93-96 (1994)
6. Zhang, B., Huirong, Y. The Internal-collision-induced Magnetic Reconnection and Turbulence (ICMART) Model of Gamma-ray Bursts. Astrophys. J. 726, 90-112 (2011)
7. Zhang, B., Mészáros, P. An Analysis of Gamma-Ray Burst Spectral Break Models. Astrophys. J. 581, 1236-1247 (2002)
8. Preece, R. D., Briggs, M. S., Mallozzi, R. S., Pendleton, G. N., Paciesas, W. S., Band, D. L. The BATSE Gamma-Ray Burst Spectral Catalog. I. High time resolution spectroscopy of bright bursts using high energy resolution data. Astrophys. J. Suppl. 126, 19-36 (2000).
9. Axelsson, M., Borgonovo, L. The width of gamma-ray burst spectra. Mon. Not. R. Astron. Soc. 447, p3150-3154 (2015)
10. Nagakura, H., Ito, H., Kiuchi, K., Yamada, S. Jet Propagations, Breakouts, and Photospheric Emissions in Collapsing Massive Progenitors of Long-duration Gamma-ray Bursts. Astrophys. J. 731, 80-97 (2011)
11. Mizuta, A., Nagataki, S., Aoi, J. Thermal Radiation from Gamma-ray Burst Jets. Astrophys. J. 732, 26-30 (2011)
12. Fan, Y., Wei, D., Zhang. F., Zhang, B. The Photospheric Radiation Model for the Prompt Emission of Gamma-Ray Bursts: Interpreting Four Observed Correlations. Astrophys. J. 755, L6-L9 (2012)
13. Ito, H. et al. Photospheric Emission from Stratified Jets. Astrophys. J. 777, 62-78 (2013)
14. Lazzati, D. Morsony, B. J., Margutti, R., Begelman, M. C. Photospheric Emission as the Dominant Radiation Mechanism in Long-duration Gamma-Ray Bursts.

Astrophys. J. 765, 103-109 (2013)

15. López-Cámara, D., Morsony, B. J., Lazzati, D. Photospheric emission from long-duration gamma-ray bursts powered by variable engines. Mon. Not. R. Astron. Soc. 442, 2202-2207 (2014)

16. Ito, H., Matsumoto, J., Nagataki, S., Warren, D. C., Barkov, M. V. Photospheric Emission from Collapsar Jets in 3D Relativistic Hydrodynamics. Astrophys. J. 814, L29-L34 (2015)

17. Lazzati, D. Monte Carlo Radiation Transfer Simulations of Photospheric Emission in Long-duration Gamma-ray Bursts. Astrophys. J. 829, 76-85 (2016)

18. Parsotan, T., Lazzati, D. A Monte Carlo Radiation Transfer Study of Photospheric Emission in Gamma-Ray Bursts. Astrophys. J. 853, 8-14 (2018)

19. Lundman, C., Pe'er, A., Ryde, F. A theory of photospheric emission from relativistic, collimated outflows. Mon. Not. R. Astron. Soc. 428, 2480-2442 (2013)

20. Rees, M. J., Mészáros, P. Dissipative Photosphere Models of Gamma-Ray Bursts and X-Ray Flashes. Astrophys. J. 628, 847-852 (2005)

21. Pe'er, A., Mészáros, P., Rees, M. J. The Observable Effects of a Photospheric Component on GRB and XRF Prompt Emission Spectrum. Astrophys. J. 642, 995-1003 (2006)

22. Vurm, I., Beloborodov, A. M. Radiative Transfer Models for Gamma-Ray Bursts. Astrophys. J. 829, 76-85 (2016)

23. Ito, H., Levinson, A., Stern, B. E., Nagataki, S. Monte Carlo simulations of relativistic radiation-mediated shocks - I. Photon-rich regime. Mon. Not. R. Astron. Soc. 474, 2828-2851 (2018)

24. Lazzati, D., Villeneuve, M., López-Cámara, D., Morsony, B. J., Perna, R. On the observed duration distribution of gamma-ray bursts from collapsars. Mon. Not. R. Astron. Soc. 436, 1867-1872 (2013)

25. Woosley, S. E., Heger, A. The Progenitor Stars of Gamma-Ray Bursts. Astrophys. J. 637, 914-921 (2006)

26. Ghirlanda, G., Nava, L., Ghisellini, G. Spectral-luminosity relation within individual Fermi gamma rays bursts. Astron. Astrophys. 511, 43-52 (2010)

27. Lu, R. et al. A Comprehensive Analysis of Fermi Gamma-Ray Burst Data. II. E p Evolution Patterns and Implications for the Observed Spectrum-Luminosity Relations. Astrophys. J. 756, 112-124 (2012)

28. Amati, L. et al. Intrinsic spectra and energetics of BeppoSAX Gamma-Ray Bursts with known redshifts. Aston. Astrophys. 390, 81-89 (2002)

29. Kocevski, D. On the Origin of High-energy Correlations in Gamma-Ray Bursts. Astrophys. J. 747, 146-155 (2012)

30. Matsumoto, J., Masada, Y., Shibata, K. Effect of Interacting Rarefaction Waves on Relativistically Hot Jets. Astrophys. J. 751, 140-157 (2012)

**Acknowledgements** This work is supported by the Grant-in-Aid for Young Scientists

(B:16K21630, 17K14308) from The Ministry of Education, Culture, Sports, Science and Technology (MEXT). Numerical computations and data analysis were carried out on XC30 and PC cluster at Center for Computational Astrophysics, National Astronomical Observatory of Japan and the Yukawa Institute Computer Facility. This work was supported in part by the Mitsubishi Foundation, a RIKEN pioneering project `Interdisciplinary Theoretical Science (iTHES)' and 'Interdisciplinary Theoretical & Mathematical Science Program (iTHEMS)'. This work was supported by NSF grant AST-1306672, DoE grant DE-SC0016369 and and NASA grant 80NSSC17K0757.

**Author Contributions** H.I. and J.M. performed numerical calculations. All the authors discussed numerical setup and the obtained results, and worked on the text of the manuscript.

**Author Information** Reprint and permission information is available at www.nature.com/reprints. The authors declare no competing financial interests. Readers are welcome to comment on the online version of this article at www.nature.com/nature. Corresponding and request for materials should be addressed to H.I. (hirotaka.ito@riken.jp).

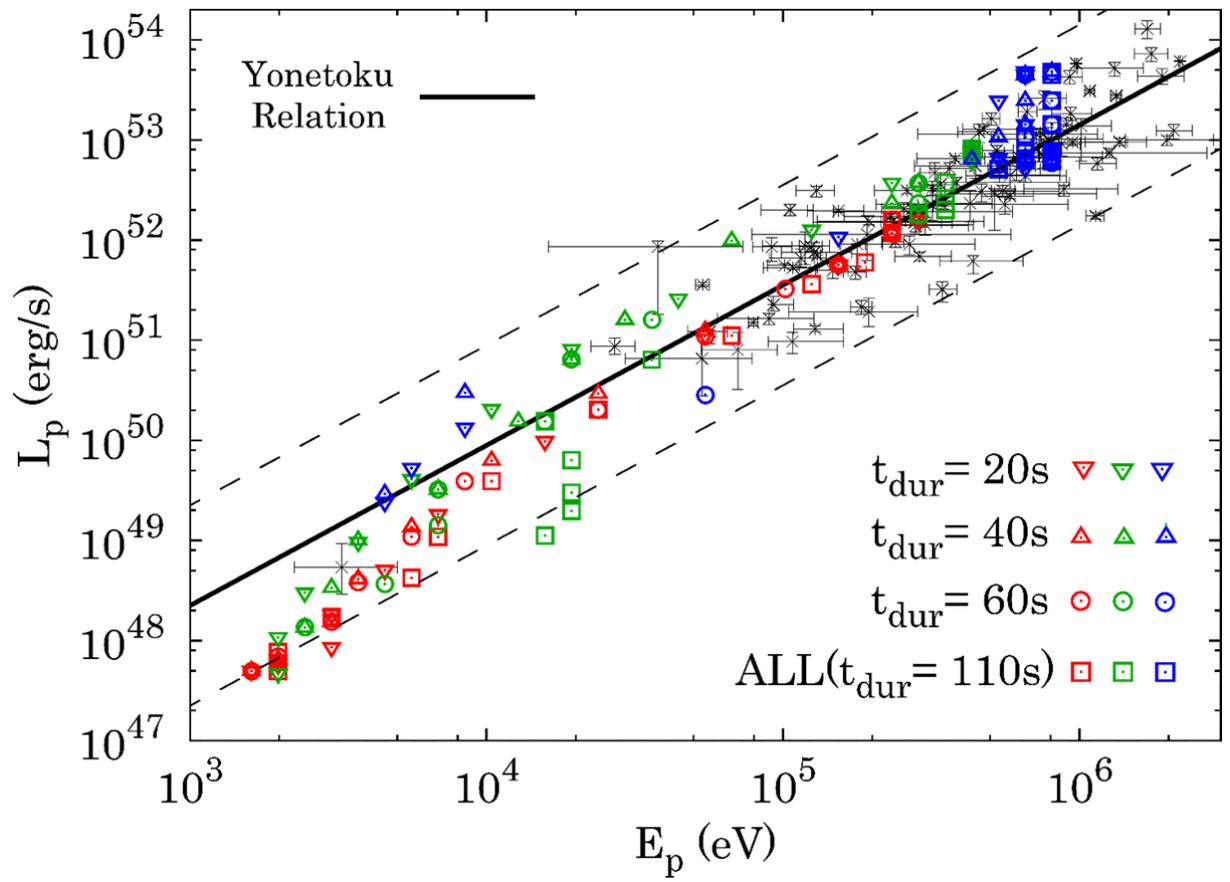

**Supplementary Figure 1| Relation between spectral peak energy $E_p$ and peak luminosity $L_p$.** Same as Figure 1, but for the simulations with a modified prescription for the evaluation of temperature.

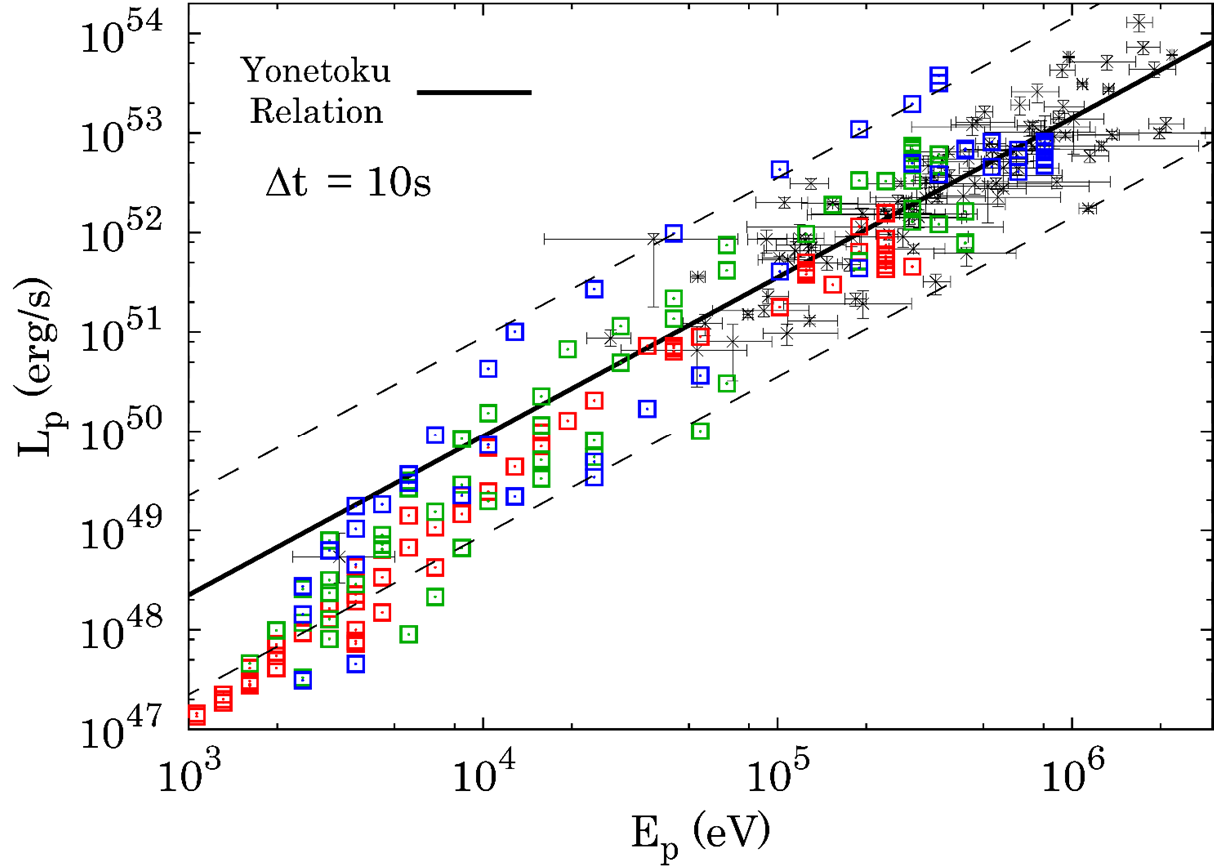

**Supplementary Figure 2| Relation between time-resolved spectral peak energy $E_p$ and peak luminosity $L_p$.** Same as Figure 1, but for a spectral peak energy and peak luminosity computed in 5 time intervals of $\Delta t$ =10 s successively taken within an observer time $t_{obs}$ = 50 s. We do not consider later observer time, since $E_p$ and $L_p$ do not change significantly thereafter.